\documentclass[aps,prl,twocolumn,amsmath,amssymb,showpacs]{revtex4-1}

\usepackage[utf8]{inputenc}
\usepackage{graphicx}
\usepackage{textcomp}
\usepackage{mathtools} % for pmatrix*
\usepackage{amsmath}
\usepackage{amssymb}
\usepackage{siunitx}
\usepackage{hyperref}
\usepackage{xcolor}
\newcommand{\ket}[1]{|#1\rangle}
\newcommand{\bra}[1]{\langle#1|}
\newcommand{\braket}[1]{\langle#1\rangle}
\newcommand{\wtilde}[1]{
  \mspace{2mu}%
  \widetilde{\mspace{-2mu}\rule{0pt}{1.5ex}\smash[t]{#1}}%
}
\newcommand{\ketbra}[2]{\ket{#1}\hspace{-0.3ex}\bra{#2}}
\newcommand{\Ham}{\mathcal{H}}
\newcommand{\dd}{\mathrm{d}}
\newcommand{\g}{\mathbf{g}}
\newcommand{\e}{\mathrm{e}}
\newcommand{\ii}{i}

\begin{document}

\title{Dynamical decoupling of spin ensembles with strong anisotropic interactions}
\author{Benjamin Merkel}
\email{B.M. and P.C.F. contributed equally to this work.}
\author{Pablo Cova Fari\~{n}a}
\email{B.M. and P.C.F. contributed equally to this work.}
\author{Andreas Reiserer}
\email{andreas.reiserer@mpq.mpg.de}

\affiliation{Quantum Networks Group, Max-Planck-Institut f\"ur Quantenoptik, Hans-Kopfermann-Strasse 1, D-85748 Garching, Germany}
\affiliation{Munich Center for Quantum Science and Technology (MCQST), Ludwig-Maximilians-Universit\"at
M\"unchen, Fakult\"at f\"ur Physik, Schellingstr. 4, D-80799 M\"unchen, Germany}

\begin{abstract} %585 of max 600 characters
Ensembles of dopants have widespread applications in quantum technology. The miniaturization of corresponding devices is however hampered by dipolar interactions that reduce the coherence at increased dopant density. We theoretically and experimentally investigate this limitation. We find that dynamical decoupling can alleviate, but not fully eliminate, the decoherence in crystals with strong anisotropic spin-spin interactions. Our findings can be generalized to all quantum systems with anisotropic g-factor used for quantum sensing, microwave-to-optical conversion, and quantum memory.

\end{abstract}
\maketitle

Distributed quantum information processing and sensing requires long-lived and efficient quantum memories \cite{afzelius_quantum_2015, wehner_quantum_2018}. A promising solid-state platform for such memories are crystals with rare-earth dopants \cite{thiel_rare-earth-doped_2011}, in which outer shell electrons shield qubits encoded in 4f-levels from  electric fields. Thus, exceptional coherence of both ground state \cite{zhong_optically_2015, rancic_coherence_2018} and optical transitions  \cite{bottger_optical_2006, merkel_coherent_2020} can be obtained. However, the small oscillator strength of the optical transitions \cite{thiel_rare-earth-doped_2011} typically requires a very large number of dopants. This hampers constructing small devices for efficient multiplexing and integration on a chip \cite{miyazono_coupling_2016, marzban_observation_2015, askarani_persistent_2020, chen_parallel_2020, weiss_erbium_2021}: When increasing the dopant concentration, the onset of dipolar interactions between spins leads to a decrease of the achievable coherence time.  While off-resonant spins can be decoupled by a series of spin-echo pulses \cite{suter_colloquium_2016}, this does not work for spins with the same Larmor frequency. Here, applying resonant pulses will change the precession frequency of each spin by altering the magnetic field generated by its neighbors. This phenomenon, termed instantaneous diffusion (ID) \cite{klauder_spectral_1962}, seems to pose a limit to the minimal size of all quantum memories and sensors based on spin ensembles. 

In this work, we theoretically and experimentally investigate to which degree this limitation can be overcome by advanced dynamical decoupling (DD) protocols. DD has developed into a powerful technique to protect the coherence of spins in solids \cite{suter_colloquium_2016}, enabling coherence time improvements by several orders of magnitude in dilute spin ensembles \cite{abobeih_one-second_2018} and sensing with unprecedented resolution \cite{degen_quantum_2017}. It has also been applied to rare-earth dopants, in particular non-Kramer's ions \cite{heinze_stopped_2013,  jobez_coherent_2015, genov_arbitrarily_2017} that experience only weak interactions. Here, we instead focus on Kramer's dopants with strong spin-spin interactions. We describe their spin state in the eigenstate basis of the Zeeman Hamiltonian
\begin{equation}
    H_Z = - \mu_B \vec{B}_0 g \vec{S}.
\end{equation}

Here, $\vec{B}_0$ is the external magnetic field, $\mu_B$ the Bohr magneton, and $\vec{S}$ the spin vector. We assume an anisotropic $g$-tensor that is diagonal in the coordinate axes $x,y,z$. Then, the spins precede around an effective magnetic field that differs from $\vec{B}_0$ \cite{car_optical_2019}. Still, the pairwise interaction between dopants can be modeled by dipolar interactions. In the Zeeman eigenbasis, the corresponding Hamiltonian reads:
\begin{equation}\label{H_dd_pauli_op}
    H_{dd} = 2 J_S \left(\hat{\sigma}_+ \hat{\sigma}_- + \hat{\sigma}_-\hat{\sigma}_+\right) + J_I \,\hat{\sigma}_z \hat{\sigma}_z.
\end{equation}

Here, $\sigma_{x,y,z}$ are the Pauli matrices, $\hat{\sigma}_\pm=\left(\hat{\sigma}_x\pm i\hat{\sigma}_y\right)/2$ the ladder operators, and we have dropped all non-secular terms like $\hat{\sigma}_+\hat{\sigma}_+$ and $\hat{\sigma}_+\hat{\sigma}_z$. The coefficients $J_S$ and $J_I$ represent the strengths of flip-flop and spectral diffusion processes, respectively. To calculate the dephasing rate, we focus on $J_I$ that describes the energy shift for a pair of two spin-1/2 dopants at a distance $r$:
\begin{equation}
    J_I = \frac{\mu_0}{4\pi\,r^3}\;\frac{h^2\gamma_\text{eff}^2}{4}\left[1-3\cos^2 \delta\right].
\end{equation}

Here, $\mu_0$ is the magnetic permeability of free space, $h$ is Planck's constant, and $\delta$ is the angle between the precession axis of the spins and the vector connecting them.  $\gamma_{\text{eff}}$ is the effective gyromagnetic ratio and depends on the projections $b_i$ of the magnetic field vector on the $g$-tensor eigenaxes as follows:
\begin{equation}   \label{g_eff}
\gamma_{\text{eff}} = \frac{\mu_B}{h} \sqrt{ \frac {g_x^4 b_x^2 + g_y^4 b_y^2 + g_z^4 b_z^2} {g_x^2b_x^2 + g_y^2b_y^2 + g_z^2b_z^2}}.
\end{equation}

When spin-lattice relaxation is negligible, one can derive a Lorentzian broadening of the transition frequency by integrating over a random distribution of dopants in the dilute ensemble \cite{geschwind_electron_1972, maryasov_spin-polarization_1982}. Its full-width-half-maximum (fwhm) linewidth is
\begin{equation} \label{eq:spinEchoLimit}
    \Delta\nu=\frac{2\pi}{9\sqrt{3}} \, \mu_0 h\gamma_\text{eff}^2 \, n_\text{eff}
.\end{equation}

Here, $n_\text{eff}$ is the concentration of spins that are flipped by the spin echo pulse \cite{agnello_instantaneous_2001}. Eq. (\ref{eq:spinEchoLimit}) implies an exponential decay of the coherence with decay constant $T_\text{SE}^{-1} = \pi \Delta\nu$, which poses a limitation on the coherence of spin ensembles with strong interactions. As an example, we now consider erbium-doped crystals, which exhibit particularly strong spin-spin interactions \cite{car_optical_2019} caused by the large effective g-factor of the erbium spins \cite{sun_magnetic_2008}.

\begin{figure*}[htb]
\includegraphics[width=2.0\columnwidth]{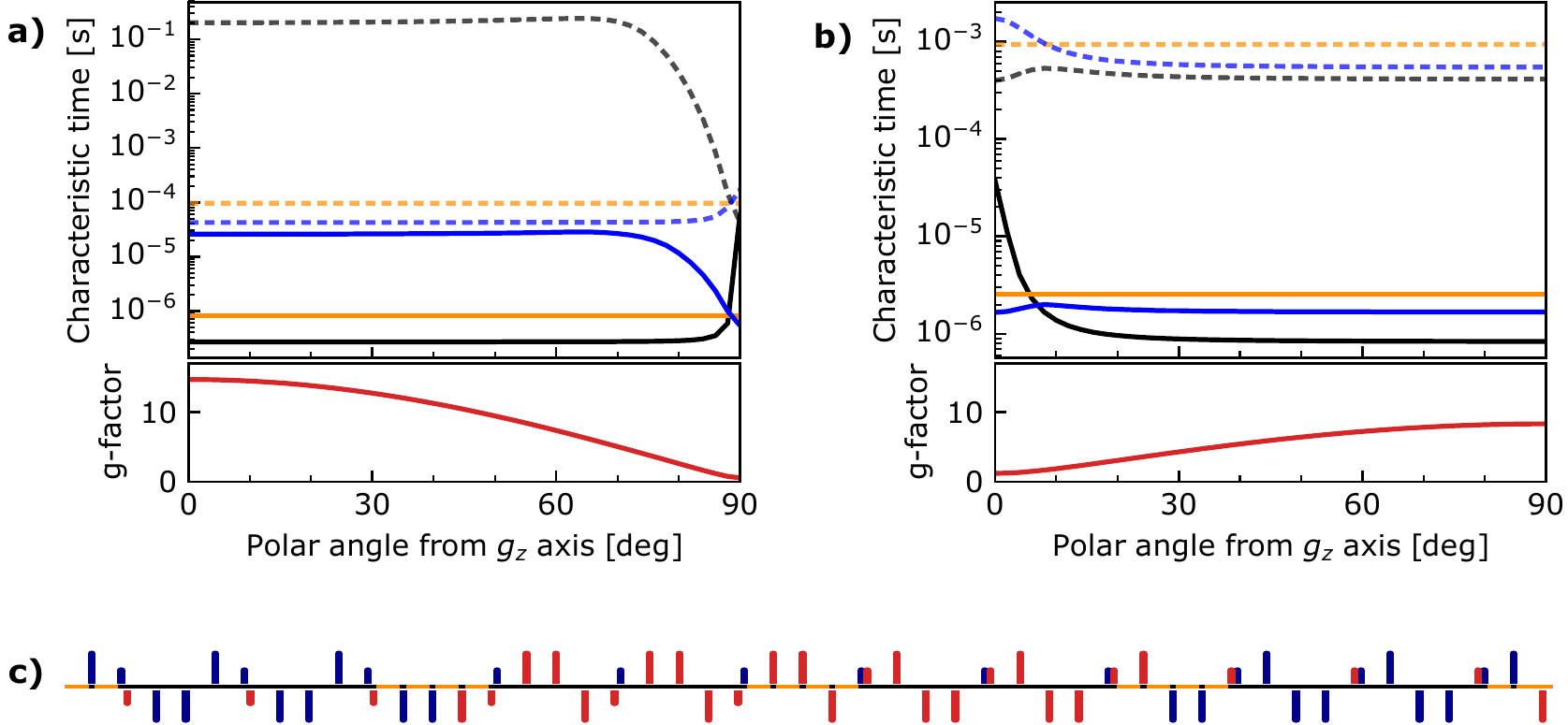}
\caption{\label{fig1:coherence_limits}
The interaction of erbium dopants at an exemplary concentration of $4\,\text{ppm}$ depends on the orientation of the external magnetic field, shown in the $g$-tensor eigenbasis for prolate (YSO site 1, panel a) and oblate tensors ($\text{CaWO}_4$, panel b). Both the flip-flop limited spin lifetime (light black dashed) and the  ID-limited spin-echo time (black solid) show an angular dependence that follows the effective $g$-factor (bottom panel). When the interactions are symmetrized using dynamical decoupling (yellow), the coherence time can be increased for magnetic field configurations with a large effective g-factor, at the price of a reduced lifetime (yellow dashed). In systems with a prolate $g$-tensor ($g_z\gg g_x,\,g_y$) further suppression of ID is possible using asymmetric decoupling sequences (blue), approaching the lifetime limit (blue dashed) at the studied concentration. In contrast, for oblate tensors ($g_z\ll g_x,\,g_y$) only a marginal further improvement of the coherence is possible. (c) Disorder-Robust-Interaction-Decoupling (DROID-60) sequence used for DD \cite{choi_robust_2020}. It consists of $\pi$- and $\pi/2$-pulses (long and short bars) along the X (blue upward), Y (red upward), -X (blue downward) and -Y (red downward) axes of the rotating frame, separated by equal spacings (black and orange lines). In the asymmetric variant, the orange delays are left out.
}
\end{figure*}

Erbium-doped crystals have recently attracted considerable interest because they enable quantum memories in the main band of optical telecommunication, which is paramount for global quantum networks. First experiments in erbium-doped yttrium-orthosilicate (Er:YSO) have demonstrated storage of optical \cite{lauritzen_telecommunication-wavelength_2010, dajczgewand_optical_2015} or microwave photons \cite{probst_anisotropic_2013} with good multiplexing capacity \cite{dajczgewand_optical_2015} and second-long hyperfine coherence at high magnetic fields, where the electronic spins are frozen to the ground state \cite{rancic_coherence_2018}. In this work, we instead study the electronic spin coherence in the low-field regime.

In Er:YSO, the Kramer's ion \cite{liu_spectroscopic_2005} erbium substitutes for yttrium in two crystallographic sites \cite{bottger_optical_2006}, each of which has two magnetically inequivalent classes. The similarity of the ionic radii of Y and Er lead to a moderate inhomogeneous broadening of the optical (few hundred MHz) and spin (few MHz) transitions. At cryogenic temperature, only the lowest crystal field level of each class is occupied. Its two-fold degeneracy is lifted in a magnetic field, and because the $g$-tensor is anisotropic, the effective g-factor depends on the magnetic field orientation, as depicted in Fig.~\ref{fig1:coherence_limits}a, bottom panel \cite{sun_magnetic_2008}. 

We consider crystals with a comparably low erbium concentration of $10\,\text{ppm}$, resulting in $n=4\,\text{ppm}$ for the even isotopes in one site. Still, the ground state lifetime is limited by flip-flop interactions \cite{lauritzen_telecommunication-wavelength_2010} (top panel, black dashed theory curve from \cite{car_optical_2019}). Following eq. (\ref{eq:spinEchoLimit}), also the spin-echo coherence time is limited by interactions (black solid line). It is shortest for the magnetic field directions that give a long lifetime. While the latter scales quadratically with $n$ \cite{car_optical_2019}, the former exhibits only linear scaling. Thus, even in the purest Er:YSO samples investigated so far, with $n\simeq0.4\,\text{ppm}$ \cite{chen_parallel_2020, merkel_coherent_2020}, the coherence of a spin echo in the ground state will be limited to the sub-millisecond range - too short for many applications.

In the following, we therefore investigate if the coherence time can be increased by DD, where a sequence of control pulses drives the spins along a path in which the interaction Hamiltonian with the environment cancels to first order \cite{brinkmann_introduction_2016}. Typical sequences employed previously eliminate frequency shifts from magnetic field inhomogeneity and off-resonant spin baths \cite{suter_colloquium_2016, heinze_stopped_2013, jobez_coherent_2015, genov_arbitrarily_2017, genov_arbitrarily_2017, kornher_sensing_2020}. The decoupling of resonant spin baths has also been achieved in spectroscopy of isotropic spin-1/2 systems \cite{waugh_approach_1968}, and anisotropic interactions have been eliminated by rotating the sample in the magnetic field. However, this technique called magic angle spinning \cite{brown_advanced_2001} is impractical for quantum memories that require a dedicated optical mode. In addition, it cannot be applied to ensembles with strong interactions, as the rotation frequency would have to be fast compared to the interaction timescale, i.e. many MHz. In such a setting, one is thus restricted to global spin (rather than sample) rotations. Then, complete decoupling of anisotropic interactions is not possible, as global spin rotations leave the isotropic ("Heisenberg") component of the interaction, $\alpha\,\hat{\vec{\sigma}}\cdot\hat{\vec{\sigma}}$, unchanged \cite{choi_dynamical_2017,choi_robust_2020, ben_attar_hamiltonian_2020}. For dipolar interactions in dilute spin baths this component reads:
\begin{equation} \label{eq:alpha}
    \alpha = \frac{\mu_0 \mu_B^2}{12\pi r^3} \sum_{i=x,y,z} g_i^2(1-3\hat{r}_i^2) = \frac{2J_S+J_I}{3}
\end{equation}

Here, $\hat{r}_i$ are the components of the unit vector connecting the spins. For an isotropic $g$-tensor with spin 1/2, the dipolar interactions between spins can be decoupled to first order as $g_x=g_y=g_z\equiv g$ and thus $\alpha \propto g^2(1-|\hat{r}|^2) = 0$. In contrast, in systems with a spin larger than 1/2, or with an anisotropic $g$-tensor, $\alpha$ will be non-zero. Thus, for general input states there is always a \textit{part} of the Hamiltonian which cannot be averaged out completely via DD in a static magnetic field \cite{choi_dynamical_2017}.

In the following we will show that under certain conditions, some coherence improvement is still possible. To this end, we first calculate the evolution of the spins caused by the microwave pulse sequence to obtain the average Hamiltonian \cite{brinkmann_introduction_2016}. For all sequences that consist only of Clifford rotations (e.g. $\pi/2$ and $\pi$) \cite{ben_attar_hamiltonian_2020, choi_robust_2020}, the average Hamiltonian keeps the form of eq.~\ref{H_dd_pauli_op}, but the coefficients change to $\tilde{J}_S$ and $\tilde{J}_I$, from which we calculate the new dephasing time constant (see supplemental material). 
We first investigate recently proposed robust pulse sequences \cite{choi_robust_2020} designed to cancel interactions with both resonant and off-resonant spin baths. Such sequences have been successfully applied to ensembles of NV-centers with \emph{isotropic} interactions \cite{zhou_quantum_2020}. Here, we instead study the case of  \emph{anisotropic} interactions. As shown in Fig.~\ref{fig1:coherence_limits}a (solid orange line), we find that a significant, but moderate improvement of the dephasing time is feasible for most magnetic field orientations, but comes at the price of a reduced flip-flop time (dashed orange line). Remarkably, the coherence becomes independent of the magnetic field angle. The reason is that the average interaction Hamiltonian for a sequence that rotates the spins in all directions for equal amounts of time reads $ H_\text{dd,iso}= \alpha\; \hat{\vec{\sigma}} \cdot \hat{\vec{\sigma}}$, which seems to suggest that an isotropic average Hamiltonian minimizes dipolar interactions in the ensemble. 

However, further improvement of the dephasing time while reducing the lifetime is possible by minimizing the component $\tilde{J}_I\,\hat{\sigma}_z\hat{\sigma}_z$. To this end, we keep the above pulse sequence, but adjust the free-evolution time periods between pulses. As detailed in the supplemental material, the weights of $(\hat{\sigma}_x \hat{\sigma}_x+\hat{\sigma}_y \hat{\sigma}_y)$ and $\hat{\sigma}_z \hat{\sigma}_z$ can be redistributed continuously \cite{choi_robust_2020,ben_attar_hamiltonian_2020}, with the  diffusion parameter ranging between $\tilde{J}_I=J_I$ (no effect) and $\tilde{J}_I=J_S$ (maximum effect).  
The latter case is shown as dashed blue line in Fig.~\ref{fig1:coherence_limits} (a). For magnetic field directions with large effective gyromagnetic ratio, the coherence is enhanced by two orders of magnitude, approaching the lifetime limit $T_2 \leq 2\cdot T_1$ at the studied concentration. However, the lifetime is strongly reduced when compared to the case without DD. While panel (a) shows the prolate $g$-tensor of YSO, which only has one large component, in (b) we instead investigate an oblate tensor, that of CaWO$_4$. Here, it turns out that asymmetric decoupling sequences do not allow for considerable improvement as compared to symmetric ones. 

After theoretically studying the improvement possible by DD, we now turn to an experimental test. We use  $0.5\,\si{\milli\metre}$ thin YSO crystals with a nominal erbium dopant concentration of $10\,\text{ppm}$, mounted in a closed-cycle cryostat at $0.8\,\si{\kelvin}$. As detailed in \cite{cova_farina_coherent_nodate}, we have established techniques to initialize, coherently control and readout the spin state of a dense ensemble of erbium dopants in a magnetic field on the order of $0.02 \,\si{\tesla}$. In particular, we achieve a $\pi$-pulse fidelity of $98\,\%$ on the ground-state spin transition, measured at the center of their inhomogeneous frequency distribution with a fwhm linewidth of $\sim9\,\si{\mega\hertz}$. When integrating over the entire line, the average flip probability is reduced by approximately one third (see supplemental material). Thus, the far-detuned parts of the ensemble will not be decoupled by our microwave pulses. As detailed in ref.~\cite{agnello_instantaneous_2001}, this reduces the effective concentration of resonant spins, but does not limit the effectiveness of DD sequences for the resonant sub-ensemble.

\begin{figure}
\includegraphics[width=1.0\columnwidth]{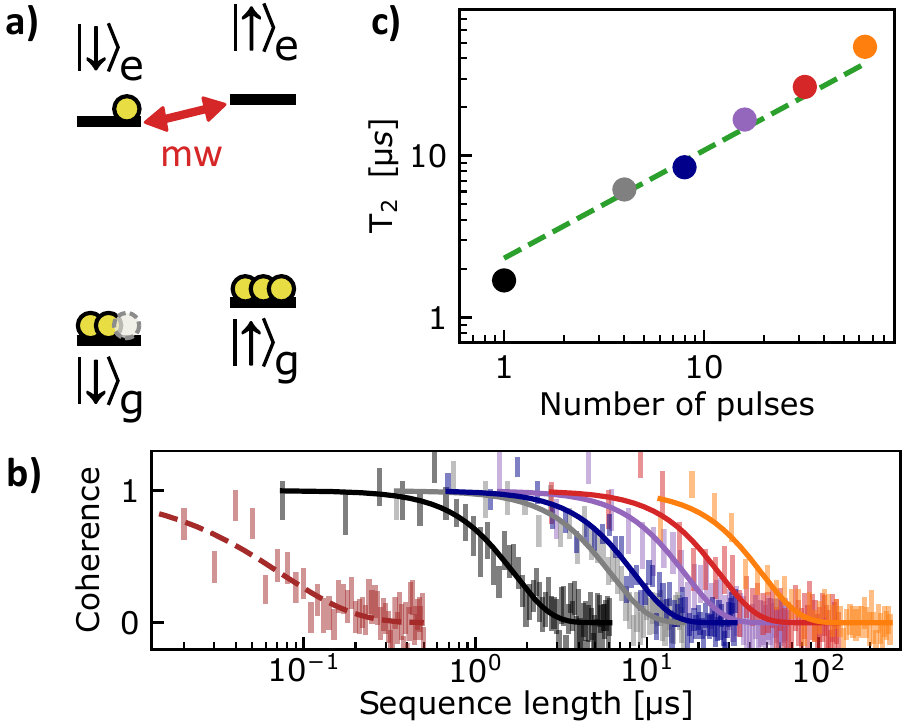}
\caption{\label{fig2:DD in excited state}
\textbf{Dynamical decoupling at a low concentration of resonant spins.} (a) The coherence of excited dopants is not limited by ID but by the interaction with the off-resonant spin baths of nuclear and ground-state electronic spins. (b) Starting from a Ramsey decay dominated by the inhomogeneous transition linewidth (brown data and exponential fit), the coherence time can be extended by XY-4 DD, leading to Gaussian decay curves $\propto\exp[-(t/T_2)^2]$ (other colors). (c) The dephasing time increases with the number of pulses as $T_2\propto(n_{\pi})^\gamma$. The green dashed line is the theoretical prediction for decoherence in a fluctuating spin bath with Lorentzian spectral noise density ($\gamma=2/3$) \cite{bar-gill_suppression_2012}.
}
\end{figure}

We first demonstrate that the quality of our microwave pulses is sufficient for DD when a robust pulse sequence is used. To this end, we apply the static magnetic field along the b axis of YSO \cite{sun_magnetic_2008}, where the magnetic classes are aligned. The microwave field is applied along D2. With a narrowband optical pulse we initialize a small fraction (around $1\,\%$) of the spin ensemble in the optically excited state  $\ket{\downarrow}_e$ (see Fig.~\ref{fig2:DD in excited state}a). We then use a microwave pulse to prepare and probe the coherence on the transition $\ket{\downarrow}_e \leftrightarrow \ket{\uparrow}_e$. As the effective g-factor $g_\text{eff,e}=10$ of the optically excited $I_{13/2}$ state differs from that of the $I_{15/2}$ ground state \cite{sun_magnetic_2008}, the concentration of resonant spins is low in this case, such that ID is negligible. The coherence of the ensemble is instead limited by \emph{off-resonant} interactions with the erbium dopants that remained in the ground state, and to a weaker extent (see supplemental material) by its interaction with the bath of Y nuclear spins \cite{kornher_sensing_2020}.

As shown in Fig.~\ref{fig2:DD in excited state}b, already a spin echo increases the coherence from an exponential decay with $0.07(1)\,\si{\micro\second}$ (Ramsey measurement, brown) to a Gaussian decay within $1.7(1)\,\si{\micro\second}$ (black). Further improvement is obtained by XY$-$4 DD (other colors). The used sequence is composed of $\pi$-pulses of $86\,\si{\nano\second}$ duration, applied alternately along the $X$ and $Y$ axes. After every four pulses, the sequence is robust against pulse imperfections \cite{suter_colloquium_2016}. As shown in (c), the improvement of the coherence with the number of $\pi$-pulses, $n_\pi$, is well fit by $T_2\propto {n_\pi}^{(2/3)}$, as predicted and previously observed for fluctuating spin baths with Lorentzian spectral noise density \cite{bar-gill_suppression_2012}. The maximally observed coherence is $48(3)\,\si{\micro\second}$ at $n_\pi=64$. At such large pulse numbers, we observe a slight broadening of the prepared optical hole that we attribute to pulse-induced heating. In spite of this observation, the obtained coherence is promising for recently proposed quantum memory protocols in the optically excited state \cite{welinski_electron_2019}, provided the excited state population is kept low enough to avoid ID.

After demonstrating that in the absence of resonant interactions DD allows for an improvement of the coherence by almost three orders of magnitude, we now turn to the ground state transition $\ket{\downarrow}_g \leftrightarrow \ket{\uparrow}_g$. Here, the concentration of resonant spins is $\sim 100$-fold higher. We start with a magnetic field orientation at $\sim130\,^{\circ}$ from the D1 axis, slightly out of the D1-D2 plane, such that the magnetic classes are detuned. Here, the effective g-factor and thus the lifetime is the largest \cite{car_optical_2019}, but ID is also the strongest. In a Ramsey measurement, we find an exponential decay of the coherence with $T_2^*=0.056(9)\,\si{\micro\second}$, caused by the inhomogeneous linewidth of the ensemble (see supplemental material). The spin echo time is $T_{\text{SE}}=0.78(8)\,\si{\micro\second}$. Compared to the spin echo in the optically excited state, we now observe an exponential rather than a Gaussian decay, and the coherence enhancement is much smaller. In addition, also DD with XY-4 and XY-8 sequences does not lead to a significant improvement: $T_{\text{XY-4}}=0.9(1)\,\si{\micro\second}$ and $T_{\text{XY-8}}=1.1(2)\,\si{\micro\second}$ (see  supplemental material). This observation is explained by the onset of ID via resonant spin-spin interactions. The observed values are in good agreement with our theoretical prediction, eq.~\ref{eq:spinEchoLimit}, which gives $T_{\text{SE}}=0.91(7)\,\si{\micro\second}$. In this calculation, as explained above, we have used a reduced effective concentration $n_{\text{eff}}=0.68(5) \cdot n$, to account for the effects of the measured inhomogeneous broadening and finite Rabi frequency (see supplemental material). 

\begin{figure}
\includegraphics[width=1.0\columnwidth]{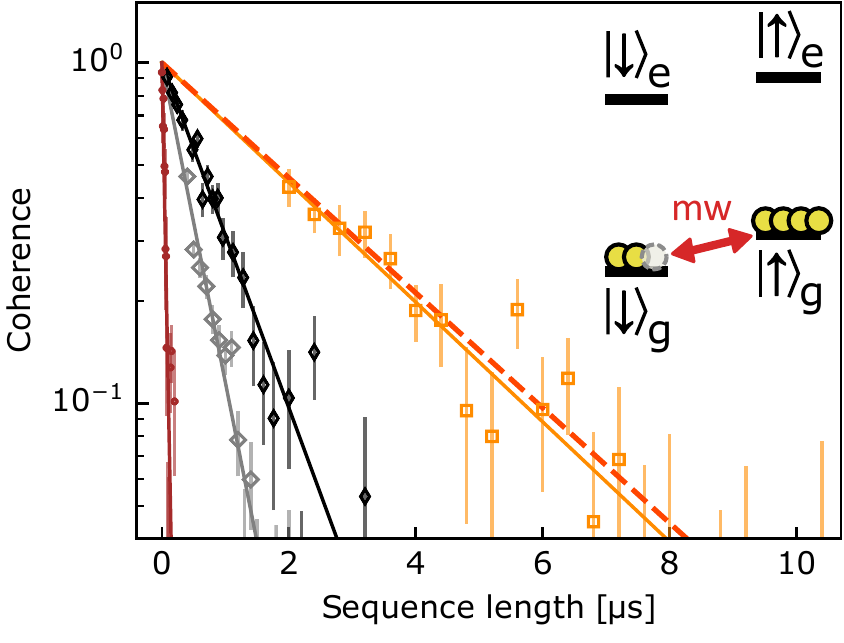}
\caption{\label{fig4:DynamicalDecoupling}
\textbf{Dynamical decoupling at high spin concentration.} On the ground-state transition (inset), more resonant spins contribute to ID, which limits the coherence. While the Ramsey measurement (brown dots and fit) gives a similar result as that in the excited state, the spin-echo decay is exponential rather than Gaussian (black diamonds and fit). Its decay constant reduces twofold when the concentration is doubled by aligning the magnetic sub-classes (grey). Application of ``DROID-60" \cite{choi_robust_2020} (yellow rectangles and fit) significantly improves the coherence time up to the theoretical limit for symmetric sequences (red dashed line).
}
\end{figure}

From this modeling, we expect that the spin echo time does not change significantly in the experimentally accessible regime of effective g-factors, where the lifetime of the spin is long enough for spin pumping \cite{car_optical_2019, cova_farina_coherent_nodate}. To demonstrate this, we next change the magnetic field direction by $40\,^\circ$, such that $g_\text{eff,g}=10.5$, similar to our previous measurements in the excited state. The field is now close to the D2 axis of YSO, but slightly tilted towards the b axis to detune the magnetic classes by $0.1\,\si{\giga\hertz}$. The spin inhomogeneous linewidth is unchanged, such that a Ramsey measurement gives $0.04(1)\,\si{\micro\second}$. Similarly, we again find an exponential decay of the spin echo with $T_{\text{SE}}=0.89(6)\,\si{\micro\second}$, see Fig.~\ref{fig4:DynamicalDecoupling} (black data and fit). This is in agreement with both our modeling and the above measurement, which proves that the dephasing is independent of the magnetic field direction over a large range.

To investigate the scaling of the decoherence with concentration, we now precisely align the magnetic field along D2, such that both magnetic classes \cite{sun_magnetic_2008} are degenerate and the density of resonant spins is increased by a factor of two. As expected, we observe a twofold decrease of the spin-echo time, finding $0.51(3)\,\si{\micro\second}$ (grey data and fit). 

The good agreement of all measurements with the theoretical prediction, the observed exponential decay and its scaling with concentration, and the ineffectiveness of XY4-DD as compared to our measurements in the optically excited state unambiguously prove that the coherence in the ground state is limited by spin-spin interactions rather than the coupling to nuclear spins or other off-resonant impurities. As explained above, these interactions cannot be eliminated by sequences that use only $\pi$-pulses. In contrast, some improvement can be expected by Disorder-Robust Interaction Decoupling, or ``DROID-60" \cite{choi_robust_2020, zhou_quantum_2020}. This sequence can decouple on-site disorder and isotropic interactions while being robust with respect to pulse imperfections, which is indispensable for DD experiments.

In our setup, we keep the magnetic field close to the D2 axis, which is the only configuration that allows for both spin initialization and fast-enough microwave Rabi frequency in our current setup \cite{cova_farina_coherent_nodate}. As shown in Fig. \ref{fig1:coherence_limits}c, we apply 48 $\pi$ and composite $\pi/2$ pulses with a duration of $33\,\si{\nano\second}$ and $19\,\si{\nano\second}$, respectively. When initializing and measuring the spins along the X-axis of the rotating frame and increasing the pulse separation, the coherence decays within $2.5(4)\, \si{\micro\second}$ (yellow data and fit). The observed enhancement matches the theoretical limitation for a symmetric pulse sequence, which according to our above analysis is around $2.66\, \si{\micro\second}$ (red dashed theory curve).

In principle, with the prolate $g$-tensor of YSO, further improvement would be possible with an asymmetric variant of the DROID-60 sequence. However, as detailed in the supplemental material, at the achievable Rabi frequency in our experiment the duration of the pulses is not short enough, such that one expects only a marginal improvement as compared to the symmetric sequence studied above.

%\section{Discussion and outlook}

To summarize, we have investigated Er:YSO as a novel platform for the study of strongly interacting spin ensembles, with applications ranging from sensing \cite{zhou_quantum_2020} to the exploration of new phases of matter \cite{choi_observation_2017}. We find that the coherence of our system is limited by interactions, but can be enhanced by a suited DD sequence that is robust with respect to pulse imperfections. Further improvement requires an increase of the Rabi frequency, e.g. via an optimized resonator geometry. This would also allow us to study the effect of DD sequences in other magnetic field configurations with a reduced effective g-factor.

We expect that our findings will be important for the improvement and integration of quantum memories and sensors. In particular, the insight that DD sequences cannot fully eliminate anisotropic spin-spin interactions, a common situation for all rare-earth dopants, quantum dots \cite{van_bree_anisotropy_2016}, and several color centers \cite{wolfowicz_qubit_2020}, has several implications: First, it seems to enforce the use of nuclear rather than electronic spins for long-lived quantum memories \cite{rancic_coherence_2018}. Second, it might stimulate research into novel materials that provide higher oscillator strength, lower inhomogeneous broadening of the optical transition, or an isotropic $g$-tensor. Third, it may guide the optimization of magnetic fields for a given combination of dopant and host. Finally, enhancing the optical depth with resonators \cite{sabooni_spectral_2013, jobez_cavity-enhanced_2014, miyazono_coupling_2016, chen_parallel_2020,  casabone_dynamic_2020, merkel_coherent_2020} seems promising to increase the coherence of rare-earth-based quantum memories by reducing the dopant concentration without sacrificing efficiency.

\section{Supplemental material}

\subsection{Magnetic interactions in systems with anisotropic g-tensors}

In this section, we will explain our mathematical modelling of the decoupling of spin-spin interactions in detail. To this end, we start with the Zeeman Hamiltonian of an effective spin-1/2 system with g-tensor $\g$ and a spin vector $\vec{S}$:
\begin{equation}\label{eq:H_Z_general}
\Ham_Z=\mu_B \vec{B}\cdot \g\cdot \vec{S}
.\end{equation}
This Hamiltonian describes the precession of the spin vector in an effective magnetic field $\vec{B}_\text{eff}=B\,\vec{b}_\text{eff}$, implicitly defined by $\vec{B}\cdot\g=\vec{B}_\text{eff}\, g_\text{eff}\,$ and $g_\text{eff}=|\vec{B}\cdot\g|/B$. An explicit expression for $\vec{b}_\text{eff}$ and $g_\text{eff}$ can be derived in the g-tensor eigenbasis, where $\g$ is diagonal with eigenvalues $g_x$, $g_y$, $g_z$. Let $b_x$, $b_y$, $b_z$ be the components of the magnetic field unit vector along the g-tensor eigenaxes. Then we can write
\begin{align}\label{eq:b_eff-def}
\frac{\vec{B}}{B}\cdot \g &= \begin{pmatrix*}[c]b_x \cr b_y \cr b_z\end{pmatrix*} \cdot
\begin{pmatrix*}[c] g_x & 0 & 0 \cr 0 & g_y & 0 \cr 0 & 0 & g_z\end{pmatrix*}
= \begin{pmatrix*}[c]b_x\, g_x \cr b_y\, g_y \cr b_z\,g_z\end{pmatrix*}
=  \vec{b}_\text{eff}\,g_\text{eff} 
,\end{align}
which gives the effective g-factor:
\begin{equation}\label{eq:g_eff-def}
g_\text{eff}=\sqrt{(b_x g_x)^2+(b_y g_y)^2+(b_z g_z)^2}
.\end{equation}

If the g-tensor is anisotropic, $g_x\neq g_y\neq g_z$, the precession axis is not aligned with the external magnetic field, $\vec{b}_\text{eff}\nparallel \vec{B}$, and the effective g-value depends on the field direction.

To calculate the eigenstates of the Zeeman Hamiltonian, we express the effective magnetic field unit vector $b_\text{eff}$ and the spin vector $\vec{S}$ by its components along the eigenaxes of the g-tensor. We choose spherical coordinates $\Theta$ and $\Phi$ to represent the direction of the effective magnetic field, and express the spin operators by Pauli matrices acting on the two-dimensional spin Hilbert space $\{\ket{+\tfrac{1}{2}}, \ket{-\tfrac{1}{2}}\}$:

%\begin{widetext}
%\begin{equation}\label{eq:H_Z_Szbasis}
%\Ham_Z = \mu_B g_\text{eff} B 
%\begin{pmatrix*}[c]\sin\Theta\cos\Phi \cr \sin\Theta\sin\Phi \cr \cos\Theta\end{pmatrix*} \cdot
%\begin{pmatrix*}[c]\hat{S}_x \cr \hat{S}_y \cr \hat{S}_z\end{pmatrix*}
%= \frac{1}{2} \mu_B g_\text{eff} B \begin{pmatrix*}[c]\cos\Theta & \sin\Theta(\cos\Phi-\ii\sin\Phi) \cr %\sin\Theta(\cos\Phi+\ii\sin\Phi) & -\cos\Theta\end{pmatrix*} 
%\end{equation}
%\end{widetext}

\begin{subequations}\label{eq:H_Z-sz_basis}
\begin{alignat}{2}
\Ham_Z =&& \mu_B g_\text{eff} B& 
\begin{pmatrix*}[c]\sin\Theta\cos\Phi \cr \sin\Theta\sin\Phi \cr \cos\Theta\end{pmatrix*} \cdot
\begin{pmatrix*}[c]\hat{S}_x \cr \hat{S}_y \cr \hat{S}_z\end{pmatrix*} 
\\ \label{eq:H_Z_Szbasis}
=&& \frac{1}{2} \mu_B g_\text{eff} B& \begin{pmatrix*}[c]\cos\Theta & \sin\Theta \,\exp(-\ii \Phi) \cr \sin\Theta \,\exp(\ii \Phi) & -\cos\Theta\end{pmatrix*} 
\end{alignat}
\end{subequations}

Diagonalizing equation~\ref{eq:H_Z_Szbasis} yields the two Zeeman eigenstates with energies $E_Z=\pm\frac{1}{2}\mu_B g_\text{eff} B$:
\begin{subequations}
\begin{alignat}{3}\label{eq:Zeeman_eigenstates}
\ket{\uparrow} =&&\cos\tfrac{\Theta}{2} \ket{+\tfrac{1}{2}}&& \;+\e^{\ii\Phi}\sin\tfrac{\Theta}{2}&\ket{-\tfrac{1}{2}} \\
\ket{\downarrow} =&&\;-\e^{-\ii\Phi}\sin\tfrac{\Theta}{2} \ket{+\tfrac{1}{2}}&& +\cos\tfrac{\Theta}{2}&\ket{-\tfrac{1}{2}}
\end{alignat}
\end{subequations}

\subsubsection*{Magnetic interaction in the Zeeman eigenbasis} \label{subsubsec:magn_moment_Zeeman_basis}

Above, we consider the g-tensor anisotropy as a tilt of the effective magnetic field axis around which the spins precede. However, this is not helpful when interacting spins are considered, because not the spins but their magnetic moments are the origin of both the Zeeman energy splitting and dipolar interactions. In systems with an anisotropic g-tensor, the spins are aligned with the tilted effective magnetic field axis, but their magnetic moments are not: they still precede around the untilted magnetic field axis as expected for any classical dipole. \cite{maryasov_bloch_2013, maryasov_anisotropic_2020}

A similar argument questions the meaning of spin operators in the context of qubit eigenstates $\ket{\uparrow}$ and $\ket{\downarrow}$: since the Zeeman Hamiltonian and the $\hat{S}_z$ operator do not commute, the spin z-component will continuously change, even though the system is in an eigenstate. To discuss processes like spin-flips and qubit rotations, it is therefore not beneficial to use the spin operators $\hat{S}_x$, $\hat{S}_y$, $\hat{S}_z$ as it was done above. 

Instead, by a more elegant choice of basis operators we can write all interactions as rotations of the effective qubit, not the spin. Such a set of qubit operators is still provided by the Pauli matrices $\hat{\sigma}_x$, $\hat{\sigma}_y$, $\hat{\sigma}_z$, similar to the spin operators, only that they are now defined in the qubit basis \cite{maryasov_spin_2012,baibekov_coherent_2014}:
\begin{subequations}
\begin{alignat}{4}
\hat{\sigma}_x=&&\ketbra{\downarrow}{\uparrow}&&+&&\ketbra{\uparrow}{\downarrow} \\
\hat{\sigma}_y=&&\;\ii\ketbra{\downarrow}{\uparrow}&&\;-&\ii&\ketbra{\uparrow}{\downarrow} \\
\hat{\sigma}_z=&&\ketbra{\uparrow}{\uparrow}&&-&&\ketbra{\downarrow}{\downarrow}
\end{alignat}
\end{subequations}

Since these operators, together with the identity matrix, form a basis of operators in the two-dimensional Hilbert space, it is possible to write any magnetic moment as a linear combination of them:
\begin{equation}\label{eq:mu_xyz_simpl}
\vec{m} = -\mu_B\, \g \cdot \vec{S} = -\frac{\mu_B}{2} \left[\vec{u}^{\,x} \hat{\sigma}_x + \vec{u}^{\,y} \hat{\sigma}_y + \vec{u}^{\,z} \hat{\sigma}_z\right]
\end{equation}

A representation of the directional vectors $\vec{u}^{\,x}$, $\vec{u}^{\,y}$, $\vec{u}^{\,z}$ in the g-tensor eigenbasis can be found by calculating $\g\cdot\hat{S}$ first in the $\hat{S}_z$-basis and then making a basis transformation to the Zeeman eigenbasis via eq.~\ref{eq:Zeeman_eigenstates}. One arrives at

%\begin{widetext}
%\begin{multline}\label{eq:mu_xyz_full}
%\vec{m}=-\frac{\mu_B}{2}\left[
%\begin{pmatrix*}[l]
%\phantom{-}g_x (\cos^2\tfrac{\Theta}{2}-\sin^2\tfrac{\Theta}{2}\cos 2\Phi) \cr 
%-g_y \sin^2\tfrac{\Theta}{2}\sin 2\Phi \cr 
%-g_z \sin\Theta \cos\Phi\end{pmatrix*} \hat{\sigma}_x
%+\begin{pmatrix*}[l]
%-g_x \sin^2\tfrac{\Theta}{2}\sin 2\Phi \cr 
%\phantom{-}g_y(\cos^2\tfrac{\Theta}{2}+\sin^2\tfrac{\Theta}{2}\%cos 2\Phi) \cr 
%-g_z \sin\Theta \sin\Phi\end{pmatrix*} \hat{\sigma}_y\right.
%\\
%+\left.\begin{pmatrix*}[l]
%g_x \sin\Theta \cos\Phi \cr 
%g_y \sin\Theta \sin\Phi \cr 
%g_z \cos\Theta
%\end{pmatrix*} \hat{\sigma}_z \right]
%\end{multline}
%\end{widetext}

\begin{equation}\label{eq:mu_xyz_full}
\begin{split}
\vec{m}=-\frac{\mu_B}{2}\left[
\begin{pmatrix*}[l]
\phantom{-}g_x (\cos^2\tfrac{\Theta}{2}-\sin^2\tfrac{\Theta}{2}\cos 2\Phi) \cr 
-g_y \sin^2\tfrac{\Theta}{2}\sin 2\Phi \cr 
-g_z \sin\Theta \cos\Phi\end{pmatrix*} \hat{\sigma}_x
\right.
\\ 
\left.
+ \begin{pmatrix*}[l]
-g_x \sin^2\tfrac{\Theta}{2}\sin 2\Phi \cr 
\phantom{-}g_y(\cos^2\tfrac{\Theta}{2}+\sin^2\tfrac{\Theta}{2}\cos 2\Phi) \cr 
-g_z \sin\Theta \sin\Phi\end{pmatrix*} \hat{\sigma}_y
\right.
\\
\left.
+ \begin{pmatrix*}[l]
g_x \sin\Theta \cos\Phi \cr 
g_y \sin\Theta \sin\Phi \cr 
g_z \cos\Theta
\end{pmatrix*} \hat{\sigma}_z \right]
\end{split}
\end{equation}

Note that because of the g-tensor anisotropy the directional vectors are not orthogonal to each other. But since they were constructed via the Zeeman eigenstates, they still fulfill $\vec{B}\cdot\vec{u}^{\,x,y}=0$ and $\vec{B}\cdot\vec{u}^{\,z}=g_\text{eff} B$, and therefore simplify the Zeeman Hamiltonian to
\begin{equation}
\Ham_Z=-\vec{B}\cdot\vec{m} = \tfrac{1}{2}\mu_B g_\text{eff} B \,\hat{\sigma}_z
\end{equation}

\subsubsection*{Dipolar interactions between resonant dopants in the Zeeman basis}

The interaction between two magnetic dipoles $\vec{m}_1$ and $\vec{m}_2$ is given by
\begin{equation}
\Ham_\text{dd}=\frac{\mu_0}{4\pi\, r^3} \left[\vec{m}_1\cdot \vec{m}_2-3(\vec{m}_1\cdot \hat{r})(\vec{m}_2\cdot\hat{r})\right]
,\end{equation}
where $\mu_0$ is the magnetic constant, $\hat{r}$ the unit vector connecting the dipoles, and $r$ their distance. For erbium dopants at the same site and class, we can express both magnetic moments by eq.~\ref{eq:mu_xyz_simpl} and rewrite the expression using ladder operators $\hat{\sigma}_\pm=(\hat{\sigma}_x \pm \ii\hat{\sigma}_y)/2$. Because in a typical experimental situation, the inhomogeneous linewidth of the spins ($\lesssim10\,\si{\mega\hertz}$) is much smaller than the Zeeman level splitting ($\sim 3\,\si{\giga\hertz}$), we can make a secular approximation and ignore terms that cause transitions between Zeeman eigenlevels of different energy, like $\hat{\sigma}_+\hat{\sigma}_z$, $\hat{\sigma}_+\hat{\sigma}_+$, etc. \cite{eden_zeeman_2014}.
We arrive at

\begin{widetext}
\begin{equation} \label{eq:H_dd_u_full}
\Ham_\text{dd}=\frac{\mu_0}{4\pi\, r^3} \left(\frac{\mu_B}{2}\right)^2\Big[
\left(\vec{u}^{\,x}\cdotp\vec{u}^{\,x}+\vec{u}^{\,y}\cdotp\vec{u}^{\,y} -3(\vec{u}^{\,x} \cdot \hat{r})^2 - 3 (\vec{u}^{\,y} \cdot \hat{r})^2 \right)\, \left(\hat{\sigma}_+ \hat{\sigma}_- + \, \hat{\sigma}_- \hat{\sigma}_+\right) + \left(\vec{u}^{\,z}\cdotp\vec{u}^{\,z} -3 (\vec{u}^{\,z} \cdot \hat{r})^2\right)\, \hat{\sigma}_z \hat{\sigma}_z \Big]
.\end{equation}
\end{widetext}

%\begin{widetext}
%\begin{multline} \label{eq:H_dd_u_full}
%\Ham_\text{dd}=\frac{\mu_0}{4\pi\, r^3} \left(\frac{\mu_B}{2}\right)^2\Big[
%\left(\vec{u}^{\,x}\cdotp\vec{u}^{\,x}+\vec{u}^{\,y}\cdotp\vec{u}^{\,y}\right)\, \left(\hat{\sigma}_+ \hat{\sigma}_- + \, \hat{\sigma}_- \hat{\sigma}_+\right) + \left(\vec{u}^{\,z}\cdotp\vec{u}^{\,z}\right)\, \hat{\sigma}_z \hat{\sigma}_z
%\\
%-3\left((\vec{u}^{\,x} \cdot \hat{r})^2 + (\vec{u}^{\,y} \cdot \hat{r})^2\right)\left(\hat{\sigma}_+\hat{\sigma}_-+\hat{\sigma}_-\hat{\sigma}_+\right)
%-3 (\vec{u}^{\,z} \cdot \hat{r})^2\, \hat{\sigma}_z\hat{\sigma}_z \Big]
%.\end{multline}
%\end{widetext}

This is a Hamiltonian of the type
\begin{equation}\label{eq:H_dd_J+-_Jz}
\Ham_\text{dd} = 2 J_S \left(\hat{\sigma}_+\hat{\sigma}_- + \hat{\sigma}_-\hat{\sigma}_+\right) + J_I \hat{\sigma}_z\hat{\sigma}_z
,\end{equation}
where we can identify $J_S$ as flip-flop coefficient and $J_I$ as spectral diffusion coefficient that induces state-dependent frequency shifts, because 
\begin{align}
&\bra{\uparrow\uparrow}\Ham_\text{dd}\ket{\uparrow\uparrow}=J_I \\
\text{and}\qquad&\bra{\uparrow\downarrow}\Ham_\text{dd}\ket{\downarrow\uparrow}=2\,J_S 
.\end{align}

\subsubsection*{Dephasing caused by dipolar interactions}

Following previous work \cite{mims_phase_1968, maryasov_spin-polarization_1982, baibekov_coherent_2014}, we can derive the inhomogeneous linewidth $\Delta\nu$, which sets the coherence time limit $T_2\leq(\pi\,\Delta\nu)^{-1}$. To this end, we average $J_S$ over a random distribution of all resonant spin-pairs. This gives:
\begin{equation}\label{eq:dipolar_linewidth}
\Delta\nu=\frac{2\pi}{9\sqrt{3}} \; \mu_0 h \gamma_\text{eff}^2 \, n_\text{eff}
\end{equation}

\begin{figure*}
\includegraphics{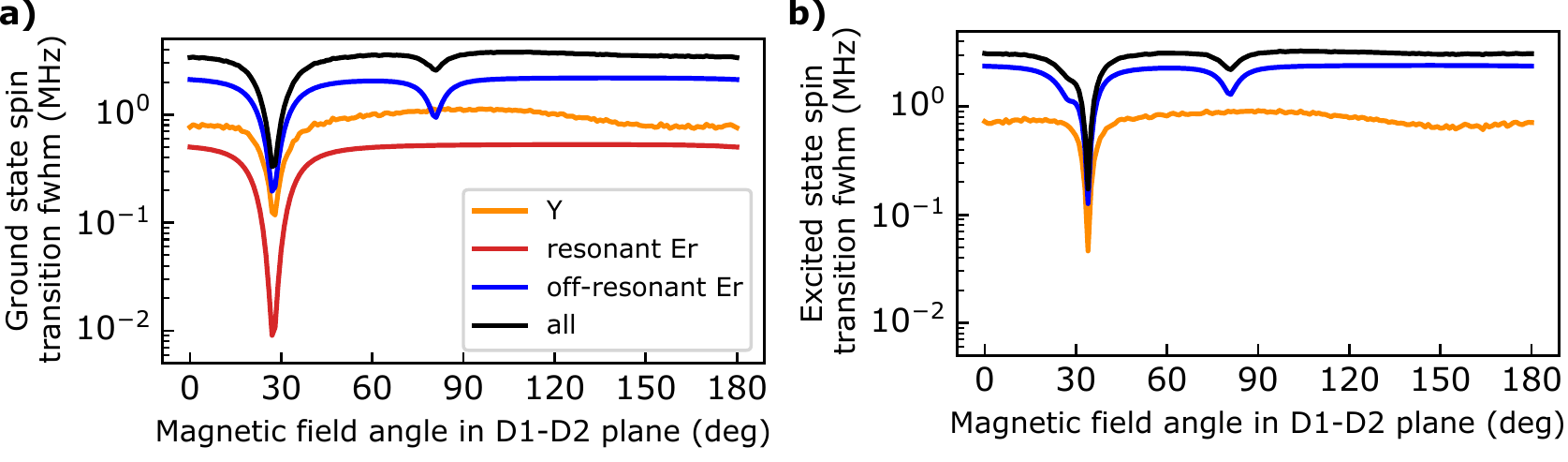}
\caption{\label{fig:diffusion_fwhm}
\textbf{Spin-bath contributions to the inhomogeneous linewidth.} The dipolar broadening of the ground state caused by resonant spins is calculated analytically using eq.~\ref{eq:dipolar_linewidth} (red). In addition, Monte Carlo simulations of 2000 bath configurations are used to calculate the inhomogeneous broadening due to the coupling to off-resonant erbium electronic (blue) and yttrium nuclear (yellow) spins. The resulting broadening depends on the magnetic field direction due to the effective g-factors of the ground (a) and excited (b) states. 
}
\end{figure*}

To also determine the influence of off-resonant erbium electronic and yttrium nuclear spins, we conduct Monte Carlo simulations. Our model uses a thermal distribution of spins at the lattice positions of Y in YSO \cite{maksimov_crystal_1970}. From this, we calculate the effective magnetic field induced by the spin bath and thus derive the resulting frequency shift for each dopant depending on the direction of the external magnetic field. Averaging over 2000 different bath configurations gives the blue and orange curves in Fig. \ref{fig:diffusion_fwhm}. It turns out that in all studied configurations, the interaction with the bath of off-resonant erbium spins gives the strongest contribution to the dephasing. This broadening can be efficiently cancelled by a simple spin-echo, as it is quasistatic (c.f. the flip-flop times in the main manuscript, or \cite{car_optical_2019}) on the timescale of the experiments. In contrast, the nuclear spin bath will show a temporal evolution on the timescale of the nuclear Larmor precession. The bulk spins precede with $2.1\,\si{\mega\hertz}/\si{\tesla}$, i.e.  $<0.05\,\si{\mega\hertz}$ at the studied fields. However, in the vicinity of the Erbium dopants, the superhyperfine interaction leads to a different precession axis and speed \cite{car_selective_2018}, with a few hundred kHz for the strongest coupled nuclei. This timescale sets a maximum pulse spacing of $\lesssim 0.3\,\si{\micro\second}$ required for effective DD \cite{kornher_sensing_2020}.

In addition to dephasing, the coupling to the nuclear spins will also lead to a slight change of the eigenstates of the electronic spin \cite{car_selective_2018}. However, as the admixture is small, we do not expect and did not observe that this has an effect on the electronic spin coherence.

\subsection{Detailed calculation of the coherence under DD}

\subsubsection*{Redistributing interaction terms by modified pulse spacings}

Engineering the average Hamiltonian in order to enhance or suppress certain interaction terms can be achieved by modifying the pulse spacings of a sequence like WAHUHA or DROID-60
\cite{farfurnik_identifying_2018, choi_robust_2020}; the resulting pulse train will inherit most robustness properties from the original decoupling sequence. For our purpose, we want to substitute the large $J_I\,\hat{\sigma}_z\hat{\sigma}_z$ terms with smaller $J_S\,\hat{\sigma}_z\hat{\sigma}_z$ ones that appear for rotated spins. To this end, we need to shorten the time intervals during which the $\hat{\sigma}_z$ component is aligned with the z-axis of the toggling frame. 

Choi et al. \cite{choi_robust_2020} already suggest a modified WAHUHA+echo sequence in which the $\hat{\sigma}_z$ component spends only a fractional time $c\,\tau$ aligned with the z-axis, compared with a time interval $(1-c)\tau/2$ spent along x and y, each. Similar modification can be made to the DROID-60 sequence; the resulting average Hamiltonian has the same form as eq.~\ref{eq:H_dd_J+-_Jz} but with new coefficients $\wtilde{J}_S$ and $\wtilde{J}_I$ in place of $J_S$ and $J_I$ \cite{choi_robust_2020}:
\begin{equation} \label{eq:mw-H_dd_Jtilde}
\Ham_\text{dd} = \wtilde{J}_S \left(\hat{\sigma}_x\hat{\sigma}_x + \hat{\sigma}_y\hat{\sigma}_y \right) + \wtilde{J}_I\,\hat{\sigma}_z\hat{\sigma}_z
,\end{equation}
with
\begin{align}
\wtilde{J}_S&=\frac{1+c}{2} \, J_S + \frac{1-c}{2} \, J_I \\
\wtilde{J}_I&= (1-c) J_S + c\, J_I
.\end{align}

Here, the parameter $c$ determines how the dipolar interaction is distributed on the spectral diffusion and the flip-flop term. For $c=1$ we retrieve the original Hamiltonian without decoupling, and for $c=1/3$ we reproduce the conventional DROID-60 sequence and the results for an isotropic Hamiltonian. Any other value of $c$ will distort the original pulse spacings and requires a more asymmetric pulse pattern. 

Remarkably, it has been shown that any DD sequence can be cast in the above form \cite{ben_attar_hamiltonian_2020, choi_dynamical_2017} as long as it consists of only Clifford operations (such as $\pi/2$ rotations around the x- or y-axis). Non-Clifford rotations could introduce other terms \cite{ben_attar_hamiltonian_2020}, for example interchanging $(\hat{\sigma}_x\hat{\sigma}_x-\hat{\sigma}_z\hat{\sigma}_z)$ with $(\hat{\sigma}_x\hat{\sigma}_z+\hat{\sigma}_z\hat{\sigma}_x)$. However, even such advanced sequences leave the isotropic part of the Hamiltonian unaffected, and we do not expect that they enable further improvement of the coherence.

\begin{figure}
\includegraphics{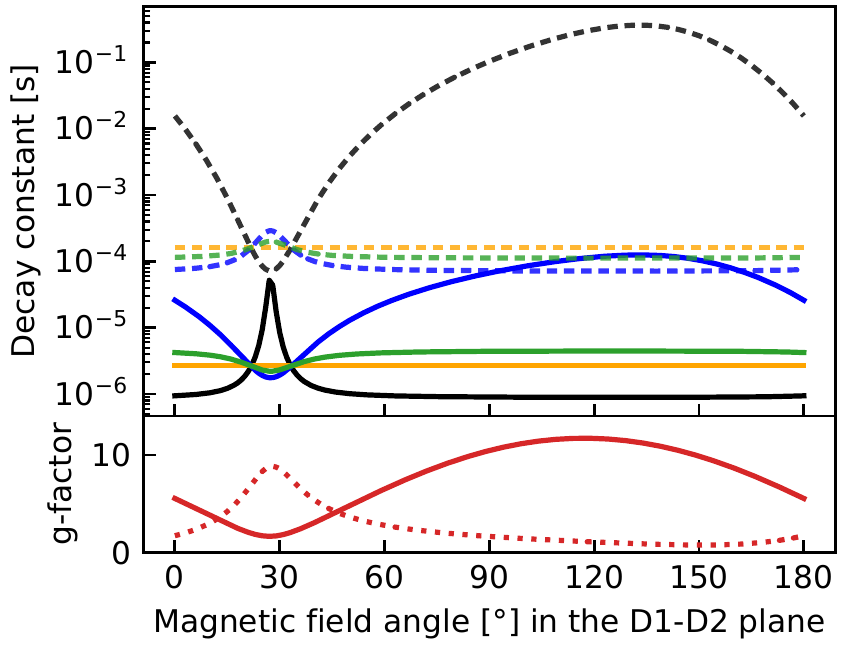}
\caption{\label{fig:real_asym_dd}
\textbf{Top: Interaction-limited lifetime and coherence of Er:YSO ground state spins.} By numerically averaging over random distributions of interacting spin pairs, we calculate the flip-flop time (dashed) and coherence time (solid) for a magnetic field in the D1-D2-plane and for various interaction asymmetry parameters $c$ (c.f.~eq.~\ref{eq:mw-H_dd_Jtilde}). Without decoupling ($c=1$, black), dipolar interactions limit the coherence time to less than $1\,\si{\micro\second}$. The improvement by symmetric decoupling sequences that yield an isotropic average Hamiltonian is only moderate ($c=0.33$, yellow), while asymmetric sequences could potentially extend the coherence time by two orders of magnitude ($c=0$, blue). However, because of a finite pulse length, the achievable asymmetry parameter was limited to $c=0.2$ (green), which performs only slightly better than symmetric sequences. 
\textbf{Bottom: effective g-factors.} The static magnetic field (here) determines the Zeeman level splitting, corresponding to an effective ground-state g-factor (solid red line), while the effect of a perpendicular microwave field (here for $\vec{B}_\text{mw}\parallel b$) is described by a microwave g-factor (dotted). 
}
\end{figure}

\subsubsection*{Optimum decoupling sequence for Er:YSO}

In YSO, for most magnetic field orientations the dipolar broadening can be reduced by substituting the large diffusion coefficient $J_I$ completely by the smaller flip-flop coefficient $J_S$. This can be realized for a value of $c=0$, resulting in an asymmetric sequence, in which several pulse spacings are contracted to zero. Such a maximally asymmetric interaction decoupling sequence produces the average Hamiltonian
\begin{equation}
\Ham_\text{dd,c=0} = \frac{J_S+J_I}{2} \left(\hat{\sigma}_x\hat{\sigma}_x + \hat{\sigma}_y\hat{\sigma}_y \right) + J_S \,\hat{\sigma}_z\hat{\sigma}_z
.\end{equation}

For this average Hamiltonian, we calculate the expected linewidth broadening and the corresponding maximum coherence time, and also the respective flip-flop rate. Fig. 1 of the main text shows this in the eigenbasis of the g-tensor. In contrast, Fig.~\ref{fig:real_asym_dd} displays the situation in the D1-D2 plane, i.e. the polarization eigenbasis of the birefringent crystal, which is easier to determine experimentally. Note that when the large coefficient $J_I$ is shifted to the flip-flop terms $(\hat{\sigma}_x\hat{\sigma}_x+\hat{\sigma}_y\hat{\sigma}_y)$, the spin lifetime is decreased (blue dashed). At the studied concentration, for an orientation of the external magnetic field between $\varphi=90^\circ$ and $160^\circ$ the coherence time can even exceed the flip-flop limited spin lifetime.

\subsubsection*{Effects of finite Rabi frequency}

The above considerations assumed instantaneous spin rotations, i.e. an infinite Rabi frequency. In a real experiment, however, very small asymmetry parameters are difficult to achieve. In our microwave resonator, we find $c\gtrsim0.2$. This lower bound is calculated by taking into account the finite pulse widths of $33\,\si{ns}$, which is not negligible compared to a typical pulse spacing of about $100\,\si{ns}$. 

As a result, the maximum coherence time extension by two orders of magnitude, expected for $c=0$ (blue in Fig. \ref{fig:real_asym_dd}), can not be achieved. With $c=0.2$ (green), the expected coherence time improvement is moderate, around $1.5$ times that expected for an isotropic Hamiltonian (yellow). The reason for this extreme sensitivity is again the high anisotropy of the g-tensor for erbium in YSO. While the interaction coefficients $J_S$ and $J_I$ are orders of magnitude apart, their mixing in the average Hamiltonian occurs only linearly, and the large $J_I$ term dominates for a wide range of parameters. 

In addition to this effect, also higher-order corrections may pose a limitation to the potential coherence time improvement, which can be estimated by calculating the commutators between the Hamiltonians for the individual free-evolution periods. While the quantitative analysis is difficult because the coefficients of the $\hat{\sigma}_i\hat{\sigma}_i$ terms are now products of $J_S$ and $J_I$ and the correct order of averaging over time and over pairs of interacting spins is not trivial, one finds that the higher-order correction favors an isotropic average Hamiltonian.

For the mentioned reasons, the use of asymmetric sequences to further enhance the coherence is not promising in our setup. It would require to drastically increase the Rabi frequency, which, however, may be possible in nanostructured samples \cite{chen_parallel_2020}.

\subsection{Calculation of the pi-pulse flip probability} \label{sec:pi_fidelity}

After initializing the spin state $\ket{\uparrow}_g$ by laser irradiation on the optical spin-flip transition $\ket{\downarrow}_g\to\ket{\uparrow}_e$ and subsequent spontaneous emission, we can measure the ground state spin polarization by probing the absorption on the $\ket{\uparrow}$ spin-preserving transition. By comparing the amplitude of this spectral antihole before and after applying a microwave $\pi$-pulse, we can determine the fidelity of spin rotations. The measured spin-flip fidelity at the center of the transition reaches $0.98$, in good agreement with the field inhomogeneity of the used MW resonator \cite{merkel_coherent_2020}. However, the average value for spins in the ensemble is lower because the finite Rabi frequency $\Omega$ leads to a decay of the flip efficiency with detuning $\Delta\nu$ from the microwave frequency \cite{agnello_instantaneous_2001}:

\begin{widetext}
\begin{equation}\label{eq:avg_pi_pulse_fidelity}
\eta=\braket{\sin^2 \theta} = \int_{-\infty}^\infty \dd\delta\nu\; \frac{1}{\pi}\frac{(\Delta\nu/2)}{(\delta\nu)^2 +(\Delta\nu/2)^2} \; \frac{\Omega^2}{\Omega^2+(2\pi\delta\nu)^2} \sin^2\left(\frac{\sqrt{\Omega^2+(2\pi\delta\nu)^2}\,t_p}{2}\right)
\end{equation}
\end{widetext}

In this formula, we take into account a Lorentzian spectral distribution of dopants with inhomogeneous fwhm linewidth $\Delta\nu$ and the excitation probability during a pulse of length $t_p$. The achievable Rabi frequency depends on the orientation of the external magnetic field and is thus different for all studied configurations:

When the field is aligned along the D2 axis of YSO, we obtained an ensemble linewidth of the spin transition of $\Delta\nu\approx 10(1)\,\si{\mega\hertz}$ in holeburning experiments and measured a Rabi frequency $\Omega=2\pi\cdot14.9(1)\,\si{\mega\hertz}$ on resonance. With these values, we calculate an average $\pi$-pulse fidelity of $\eta\approx0.78(6)$ for a $\pi$-pulse length of $33\,\si{\nano\second}$. 

When we orient the static magnetic field at an angle $\sim130^\circ$ with the D1-axis, the $\pi$-pulse fidelity drops to about $0.68(5)$. Finally, for the presented experiments in the optically excited state, the static magnetic field was oriented along b and the microwave magnetic field along D2. In this case we measure a Rabi frequency of $2\pi\cdot6.2(2)\,\si{\mega\hertz}$ and calculate a $\pi$-pulse fidelity estimate of $0.65(5)$.

\subsection{XY-Decoupling on the ground state transition}

In experiments on the optically excited state transition, we achieved a coherence time extension by almost three orders of magnitude when increasing the number of pulses in an XY-4 decoupling sequence. The observed value and Gaussian decay are in excellent agreement with the expectation from the coupling to a fluctuating bath of Yttrium nuclear spins \cite{kornher_sensing_2020}. 

\begin{figure}
\includegraphics{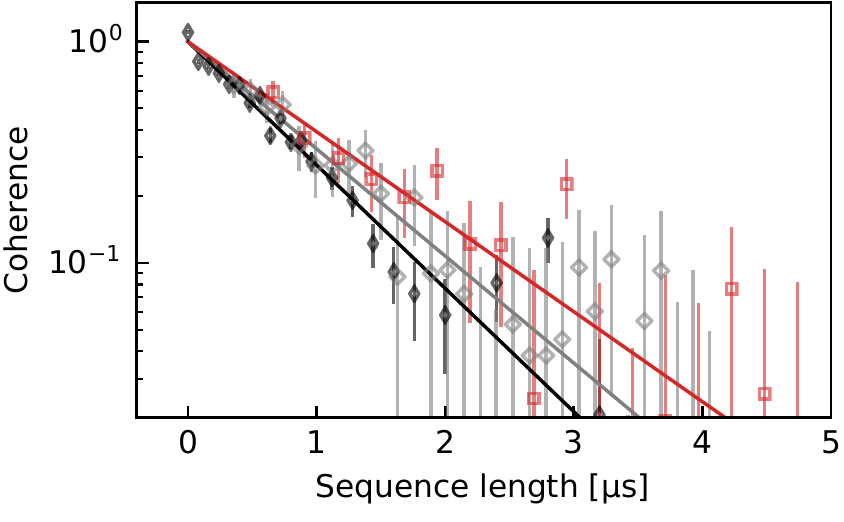}
\caption{\label{fig:dd_130deg}
\textbf{Ground-state decoupling with XY sequences.} For the magnetic field oriented at $\varphi=130^\circ$ from the D1-axis, we measure a spin-echo time of $0.78(8)\,\si{\micro\second}$ (black data with exponential fit). As expected from our theoretical modeling, with XY-4 (grey) and XY-8 (red) decoupling sequences we achieve only a moderate coherence time extension to $0.9(1)\,\si{\micro\second}$ and $1.1(2)\,\si{\micro\second}$, respectively, which proves that the coherence is limited by dipolar interactions with a resonant spin bath. 
}
\end{figure}

In contrast, on the ground state spin transition, we observed exponential decays of the spin echo at all studied concentrations and magnetic field orientations, as described in the main paper. In figure~\ref{fig:dd_130deg} we show additional measurements, in which we investigate the coherence in DD experiments using XY-4 (grey) or XY-8 (red) sequences. As expected, we observe only a slight increase in the coherence time as compared to the spin echo (black), which we attribute to the finite bandwidth of the pulses that reduces the effective concentration, as detailed in Sec. \ref{sec:pi_fidelity}. The shown measurements were performed at a magnetic field orientation of $\varphi=130^\circ$ from the D1-axis, where the microwave g-factor limits the maximum Rabi frequency to $2\pi\cdot6.4(5)\,\si{\mega\hertz}$, which results in an effective flip probability of only $0.65(5)\,\%$.

\section{acknowledgements}

We acknowledge discussions with Nir Bar-Gill and Friedemann Reinhard. This project received funding from the European Research Council (ERC) under the European Union's Horizon 2020 research and innovation programme (grant agreement No 757772), from the Deutsche Forschungsgemeinschaft (DFG, German Research Foundation) under Germany's Excellence Strategy - EXC-2111 - 390814868.

% Currently 2942 (built-in word count) of max 3750 words.
% 6 inline equations each corresponding to 16 words: add 96
% about 120 words per figure, i.e. 480 on total (two-column).
% So should be within limits

\bibliography{bibliography.bib}

\end{document}